\documentclass{article}
\begin{document}

\centerline{\bf \huge The importance of the Ising model}

\vspace{,2in}

\centerline{\bf \Large Barry M. McCoy$^1$  and Jean-Marie Maillard$^2$}

\vspace{.1in}
\centerline{\bf 1. State University of New York, Stony Brook, NY}
\centerline{\bf 2.  University Paris 6, UMR 7600 CNRS, Paris, France}

\vspace{.2in}

\centerline{\bf Abstract}

\vspace{.1in}

Understanding the relationship which integrable (solvable) models, 
all of which  possess very special symmetry properties, have with the
generic non-integrable models that are used to describe real
experiments, which do not have the symmetry properties, is one of the
most fundamental open questions in both statistical mechanics and
quantum field theory.  
The importance of the two-dimensional Ising model in a magnetic field
is that it is  the simplest
system where this relationship may be concretely studied. We here review
the advances made in this study, and concentrate on the magnetic 
susceptibility which has revealed an unexpected natural boundary
phenomenon. When this is
combined with the Fermionic representations of conformal
characters, it is suggested that the scaling theory, 
which smoothly connects the lattice with the correlation length scale, 
may be incomplete for $H \neq 0$.

\vspace{.1in}

\section{Introduction}

It may be rightly said that the two dimensional Ising model for $H=0$ 
is one of the most important systems studied in theoretical physics. 
It is the first statistical mechanical 
system which can be exactly solved which exhibits
a phase transition. From the exact results for the free
energy~\cite{ons}, spontaneous magnetization~\cite{ons2},\cite{yang} 
and correlation functions~\cite{ko}-\cite{jm} a point of
view has been developed, which embraces the concepts of scaling, 
universality and conformal field theory, that extends the exact results of
the Ising model to more general situations. These concepts are  widely
used to analyze both experiments and models of critical
phenomena. Furthermore the correlation functions provide very concrete
realizations of the concepts of mass and wave function renormalization
used to define Euclidean quantum field theories.
  
However, starting with the work of Nickel~\cite{nic1},\cite{nic2} 
on the magnetic
susceptibility new properties of the Ising model have been 
uncovered~\cite{ongp}-\cite{mccoy21} which go beyond what has been 
seen in the computations of the free
energy, spontaneous magnetization and correlation functions. These new
features need to be explored to see if there is relevant physics which
is not incorporated in our current view of critical phenomena. In this
article we will review these new phenomena and the relation they have with
scaling theory and Euclidean quantum field theory.  

In section 2 we  define what will be meant by an Ising
model. In section 3 we review the known exact results for $H=0$. 
In addition to the well known results for the free energy~\cite{ons}
and the magnetization~\cite{ons2},~\cite{yang} we
will put particular emphasis on the magnetic susceptibility which has
an expansion analogous \cite{wmtb},\cite{nic1},\cite{nic2} 
to a Feynman diagram expansion.
These Ising model integrals share with Feynman diagram integrals 
the property that   
the integrands are algebraic functions. It was shown, long ago, by
Kashiwara and Kawai~\cite{kk}, that these Feynman integrals 
are holonomic (i.e. 
they satisfy overdetermined systems 
of partial differential equations).
However, the infinite sum of diagrams will 
not have this property\footnote[1]{It has first remarked by Enting 
and Guttmann~\cite{Enting}, on the anisotropic lattice,
that the susceptibility of the square Ising model 
is not D-finite~\cite{Enting}. The 
isotropic susceptibility is also non-holonomic (but this is not a consequence 
of the previous anisotropic result).}. 
These problems of power series expansions, where the infinite sum has
different analytic  properties than the coefficients of any finite
power of the expansion parameter, are seen in the magnetic susceptibility
of the Ising model~\cite{Enting}. When expanded about
 $T=\, T_c$, this expansion of the 
susceptibility is not convergent but is only asymptotic \cite{ongp}\cite{cgnp}. This 
feature also occurs in Quantum Electrodynamics~\cite{dyson} and
Quantum Chromodynamics~\cite{thooft}.  

In section 4 we extend our considerations to $H\neq 0$, where much less
is known and there is much to be learned. We will present perturbative studies, 
for $H$ near zero, of the two-point function~\cite{mccoy52},
and the two-particle scattering amplitude~\cite{zz}. We will also
present integrable perturbations of conformal field theory~\cite{bpz}-\cite{zam2}  
about $T=T_c,H\neq 0$, about the Lee-Yang edge~\cite{ly1},~\cite{ly2},
 and the relation to Rogers-Ramanujan identities~\cite{mccoy50}-\cite{wp}. 
We will finally discuss scaling 
theory~\cite{ongp},~\cite{cgnp},~\cite{af},~\cite{caselle},
 and the study  of the general non integrable perturbation of~\cite{fz}.   
 
We conclude, in section 5, with an answer to the question of ``Why is the
Ising model is important?''

\section{What is the Ising model?}

We begin by defining what we mean by the two-dimensional Ising model
in a magnetic field.

The simplest, and most well known, case is for nearest neighbor
 interactions on a
square lattice defined by the classical interaction energy
\begin{equation}
\hspace{-0.01in}{\mathcal E}_I(H)\,=\,\, -\sum_{j=-L^v}^{L^v}\sum_{k=-L^h}^{L^h}
\{E^v \cdot \,\sigma_{j,k}  \; \sigma_{j+1,k}
+E^h \cdot\,\sigma_{j,k} \; \sigma_{j,k+1}\,+ H \cdot \,\sigma_{j,k}\},
\label{nni}
\end{equation}
where $\sigma_{j,k}=\pm 1$ specify the ``spin'' at row $j$ and
column $k$ of a square lattice. This definition can be extended to
nearest neighbor interactions on other lattices by a straightforward
expansion of the notation. We will impose either periodic, or cylindrical,
boundary conditions.

For this interaction energy we are interested in computing the
partition function
\begin{equation} 
Z_{L^v,L^h}(T,H)\,\,\, =\, \,\,\, \sum_{\sigma=\pm}\exp (-{\mathcal E}_I(H)/k_BT), 
\label{part}
\end{equation}
where $k_B$ is the Boltzmann constant and the sum is over all values
of the spins $\sigma_{j,k}$. 

From the partition function (\ref{part}) one gets
the free energy $F(T,H)$ in the thermodynamic limit 
\begin{equation}
-F(T,H)/k_BT \,\,  = \,\,  \lim_{L^v,L^h\rightarrow \infty}\frac{1}{L^vL^h}\, 
\ln Z_{L^v,L^h}(T,H),
\label{free}
\end{equation}
the magnetization 
\begin{equation}
\hspace{-0.2in}{\mathcal M}(T,H)\, =\,\, -\frac{\partial}{\partial  H}F(T,H)
\, =\,\, \langle \sigma_{0,0}\rangle,
\label{mag}
\end{equation}
the magnetic susceptibility
\begin{equation}
\hspace{-0.02in}\chi(T,H)\, =\,\, \frac{\partial M(T,H)}{\partial  H}
\, =\,\, \frac{1}{k_BT}\sum_{M=-\infty}^{\infty}\sum_{N=-\infty}^{\infty}
\{\langle \sigma_{0,0}\sigma_{M,N}\rangle
 -{\mathcal M}^2(T,H)\}, 
\label{sus}
\end{equation}
and the internal energy
\begin{equation}
u\,\,=\,\,\, k_B T^2\,\frac{\partial}{\partial T}F(T,H)/k_BT, 
\label{udef}
\end{equation}
where we have used the definition of the
thermal average of an operator $O$ 
\begin{equation}
\langle O \rangle\,=\,\,\lim_{L^v,L^h\rightarrow
  \infty}Z^{-1}_{L^v,L^h}(T,H)\cdot \,\sum_{\sigma=\pm}
O\exp(-{\cal E}_I(H)/k_BT).
\label{corr}
\end{equation}
We note, in particular, that the magnetic 
susceptibility, at $H=0$, is written, in
terms of the two-point function, as
\begin{equation}
k_BT\cdot \, \chi(T,0)\,\, = \,\, \sum_{M=-\infty}^{\infty}\sum_{N=-\infty}^{\infty}
\{\langle \sigma_{0,0}\sigma_{M,N}\rangle-{\mathcal M}^2\},
\end{equation}
where
\begin{equation}
{\mathcal M}\,\, =\,\, \, \lim_{H\rightarrow 0+}M(T,H), 
\end{equation}
is the spontaneous magnetization.
For the nearest neighbor interaction (\ref{nni}), the internal energy
(\ref{udef}) reads:
\begin{equation}
u\,=\,\,\,  -E^v \cdot \, \langle \sigma_{0,0}\sigma_{1,0}\rangle \,
\, -E^h\cdot \, \langle \sigma_{0,0}\sigma_{0,1}\rangle \, 
\, -H \cdot \,\langle \sigma_{0,0}\rangle.
\label{nnu}
\end{equation}

The interaction energy of nearest neighbor Ising model (\ref{nni}) 
may be generalized to
interactions farther than nearest neighbors with interaction energy: 
\begin{eqnarray}
{\mathcal E}_G&=&-\sum_{j=-L^v}^{L^v}\sum_{k=-L^h}^{L^h}
\sum_{j'=-L^v}^{L^v}\sum_{k'=-L^h}^{L^h}
E(|j-j'|,|k-k'|)\cdot \, \sigma_{j,k}\sigma_{j',k'}
\nonumber\\
&&\quad -H \cdot \, \sum_{j=-L^v}^{L^v}\sum_{k=-L^h}^{L^h}\sigma_{j,k}.
\label{geni}
\end{eqnarray}
This reduces to (\ref{nni}) when $E(1,0)=E^v/2,~E(0,1)=E^h/2$ and all
other $E(j,k)=0$. The free energy, magnetization and susceptibility are 
all computed by replacing ${\mathcal E}_I$ in (\ref{part}) by
${\mathcal E}_G$. The internal energy (\ref{nnu}) generalizes to
\begin{equation}
u\,\,=\,\,\, -\sum_{j=-L^v}^{L^v}\sum_{k=-L^h}^{L^h}E(|j|,|k|) \cdot \, 
\langle\sigma_{0,0}\sigma_{j,k}\rangle 
\,\, -H \cdot \, \langle \sigma_{0,0}\rangle.
\label{genu}
\end{equation}

\section{What do we know for $H=0$ ?}
 
We have obtained more exact results for the nearest neighbor Ising 
model at $H=0$ than for any other system in statistical mechanics.

\subsection{Free energy}
 
The free energy of the nearest neighbor Ising model with $H=0$ was
computed by Onsager~\cite{ons} in 1944
\begin{eqnarray}
\hspace{-0.3in}&&-F/k_BT\,\, \, = \,\, \, \, \ln 2 \\
\hspace{-0.3in}&&+\frac{1}{8\pi^2} \int_{0}^{2\pi} d\theta_1\int_{0}^{2\pi}d\theta_2 
\ln \Big[\cosh \frac{2E^h}{k_BT}\cosh\frac{2E^v}{k_BT}
-\sinh\frac{2E^h}{k_BT}\cos\theta_1
-\sinh\frac{2E^v}{k_BT} \cos\theta_2 \Big]. 
\nonumber
\label{isingfree}
\end{eqnarray}
This free energy has a singularity at $T=T_c$ determined
by
\begin{equation}
\sinh 2E^v/k_BT_c \cdot \, \sinh 2E^h/k_BT_c \,\, \, = \, \, \, 1.
\end{equation}
This temperature $T_c$ is referred to as the critical temperature.
The free energy may be expanded, about $T=\,T_c$, as
\begin{equation}
\hspace{-0.3in}-F/k_BT\,\, = \,\,\, 
(T-T_c)^2 \cdot \, \ln|T-T_c| \cdot \, F_1(T-T_c) \, \,\, +F_2(T-T_c), 
\label{freeexpansion}
\end{equation}
where $F_1$, and $F_2$, are analytic at $T=T_c$ (i.e. they both have convergent
power series expansions).

\subsection{Spontaneous magnetization}

The spontaneous magnetization was first announced by 
Onsager~\cite{ons2} in 1949,
and a proof was given by Yang~\cite{yang} in 1952
\begin{equation}
{\mathcal M}\,\, =\, \,\,\, (1-k^2)^{1/8}, 
\label{spont}
\end{equation}
where
\begin{equation}
k\,\, =\,\, \, (\sinh 2E^v/k_BT \cdot \,\sinh 2E^h/k_BT)^{-1}, 
\end{equation}
for $T<T_c$, and is zero for $T>T_c$.
The history of the Onsager result is given in a most interesting paper
of Baxter~\cite{bax}. 

\subsection{Correlation functions}

All correlation functions of the nearest neighbor Ising model at $H=0$
may be expressed as determinants. These are particularly simple for
the row correlation $\langle \sigma_{0,0}\sigma_{0,N}\rangle$, and the diagonal
correlations $\langle \sigma_{0,0}\sigma_{N,N}\rangle$, which are given by
\begin{equation}
D_N\,\, = \,\,  \,  \,
\begin{array}{|llll|}
{a}_0&{a}_{-1}&\cdots&{a}_{-N+1}\\
{a}_1&{ a}_0&\cdots&{a}_{-N+2}\\
\vdots&\vdots&&\vdots\\
{a}_{N-1}&{a}_{N-2}&\cdots&{a}_0
\end{array}
\label{detdn}
\end{equation}
with
\begin{equation}
a_n\, \, =\,\, \,  {1\over 2\pi}\int_{0}^{2\pi}\, d\theta \cdot \, 
e^{-in\theta} \cdot \, 
\left[ {(1-\alpha_1e^{i\theta})(1-\alpha_2e^{-i\theta})\over
(1-\alpha_1e^{-i\theta})(1-\alpha_2e^{i\theta})}\right]^{1/2}, 
\label{dn2}
\end{equation}
where for $\langle\sigma_{0,0}\sigma_{N,N}\rangle$
\begin{equation}
\hspace{0.02in}\alpha_1\,=\,\, 0, \quad \quad \quad 
\alpha_2\,= \,\, (\sinh 2E^v/k_BT \sinh2E^h/k_BT)^{-1}, 
\label{dn3}
\end{equation}
and for $\langle \sigma_{0,0}\sigma_{0,N}\rangle$
\begin{equation}
\hspace{0.1in}\alpha_1= \, e^{-2E^v/k_BT}\cdot \,\tanh E^h/k_BT,
 \quad \,\,\,
\alpha_2 = \, e^{-2E^v/k_BT} \cdot \,\coth E^h/k_BT.
\label{dn4}
\end{equation}

From this determinantal representation (\ref{detdn}) of $\langle
\sigma_{0,0}\sigma_{0,N}\rangle$, and $\langle
\sigma_{0,0}\sigma_{N,N}\rangle$, it is easy to obtain 
the behavior of the correlation function as $T\rightarrow T_c$.
The integrals $a_n$ all have logarithmic singularities at $T= \, T_c$.
and we find~\cite{ongp},~\cite{ayp} for both the row 
and the diagonal correlations that $D_N$ has an expansion of the form
\begin{equation}
 D_N\, \, = \, \,  \, \, \, \sum_{p=0}^{N}\sum_{n=0}^{\infty} \, \, 
 d^{(p,n)}(\ln |T-T_c|)^p \cdot \, |T-T_c|^{p^2+n}, 
\label{short}
\end{equation}
where for each $p$ the sum over $n$ converges (and thus defines an
analytic function).
The correlations 
$\langle \sigma_{0,0}\sigma_{M,N}\rangle$ 
have a similar expansion in terms of powers of $\ln |T-T_c|$ 
multiplied by functions which are analytic at $T=T_c$.

When $T=T_c$ the determinant for the diagonal correlation reduces to a
Cauchy determinant. It is explicitly evaluated as~\cite{wu}, yielding
\begin{equation}
\langle \sigma_{0,0}\sigma_{N,N}\rangle
\, \,  = \,  \, \, \Bigl(\frac{2}{\pi}\Bigr)^N \cdot \,
\prod_{j=1}^{N-1}\left(1-\frac{1}{4j^2}\right)^{l-N}.
\label{diagexplicit}
\end{equation}
From this, we find as $N\rightarrow  \infty$
that
\begin{equation}
\langle\sigma_{0,0}\sigma_{N,N}\rangle
 \,\,= \, \,\,  \frac{A_c}{N^{1/4}} \cdot \,\left(1\, -\frac{1}{64N^2}+O(N^{-4})\right), 
\label{corrtc}
\end{equation}
with
\begin{equation}
A_c\, \, =\, \, \, 2^{1/12} \cdot \, \exp[3\zeta'(-1)]\, \sim
 \, \, \, 0.6450\, \, \cdots
\label{ac}
\end{equation}
where $\zeta'(z)$ is the derivative of the zeta function.

However, for $T\neq T_c$, the representation (\ref{detdn}) is not an efficient 
way to study the limit $M,N\rightarrow \infty$.

\subsection{Form factor expansions}

To study the correlation functions $\langle
\sigma_{0,0}\sigma_{M,N}\rangle$ when $M,N\rightarrow \infty$,
the determinants  are re-expressed in
an ``exponential form''~\cite{wmtb} for $T<T_c$ as
\begin{equation}
\langle \sigma_{0,0}\sigma_{M,N}\rangle \, \,  = \, \,  \,\, {\mathcal M}^2
\cdot \, \exp\sum_{n=1}^{\infty}F^{(n)}_{-}(T;M,N),
\label{expm}
\end{equation}
and, for $T>T_c$
\begin{equation}
\hspace{-0.02in}\langle \sigma_{0,0}\sigma_{M,N}\rangle
\, \,   =\,\,\,  {\mathcal M}_{+}^2 \cdot \, 
\sum_{j=1}^{\infty}\, G^{(2j-1)}(T;M,N) \cdot \, \exp\sum_{n=1}^{\infty}F^{(2n)}_{+}(T;M,N),
\label{expp}
\end{equation}
where 
\begin{equation}
{\mathcal M}_{+}\, =\,\, \, \, [1 \,-(\sinh 2E^v/k_BT \cdot  \,\sinh 2E^h/k_BT)^2]^{1/8}, 
\end{equation} 
is referred to as the disorder parameter and is the value of the 
spontaneous magnetization on
the dual lattice where $E^v$
and $E^h$ are replaced by $E^{*h}$ and $E^{*v}$ defined as  
\begin{equation}
\sinh 2E^{*v}/k_BT \,
= \, \, 1/\sinh 2E^{v}/k_BT, \quad  \, \,  
 \, \sinh 2E^{*h}/k_BT \,=\, \,1/\sinh 2E^{h}/k_BT.
\label{dual}
\end{equation}
The functions
$F^{(n)}_{\pm}(T;M,N)$ and $G^{(n)}(T;M,N)$ are $n$-dimensional
  integrals. These exponentials can be expanded, and 
terms combined~\cite{wmtb},~\cite{nic1},~\cite{nic2} into 
what is referred to as the
  form factor expansion. For $T<T_c$ this expansion is
\begin{equation}
\langle\sigma_{0,0}\sigma_{M,N}\rangle
\,\, =\, \, \, (1-t)^{1/4} \cdot \, \{1+\sum_{n=1}^{\infty}f^{(2n)}_{M,N}(T)\}, 
\label{ffm}
\end{equation}
with $f^{(2n)}_{M,N}(T)$ a $2n$-dimensional integral and:
\begin{equation}
t\,\, =\,\, \, \, (\sinh 2E^v/k_BT \cdot \, \sinh 2E^h/k_BT)^{-2}.
\end{equation}
For $T>T_c$ the form factor expansion reads
\begin{equation}
\langle\sigma_{0,0}\sigma_{M,N}\rangle
\,\,  =\,\, \,  \,  (1-t)^{1/4} \cdot \, \sum_{n=0}^{\infty}f^{(2n+1)}_{M,N}(T), 
\label{ffp}
\end{equation}
where $f^{(2n+1)}_{M,N}(T)$ is a $\, 2n+1$ dimensional integral, and
\begin{equation}
t\,\, =\, \,\, (\sinh 2E^v/k_BT \cdot \, \sinh 2E^h/k_BT)^{2}.
\end{equation}
For the diagonal correlations $\langle
\sigma_{0,0}\sigma_{N,N}\rangle$
a simpler alternative form of $f^{(n)}_{N,N}(t)$ is given in~\cite{mccoy2}, and
proven in~\cite{mccoy3}.

The behavior of $\langle \sigma_{0,0}\sigma_{M,N}\rangle$ is easily
obtained, from this form factor expansion when $T \neq T_c$, because
$f^{(n)}_{M,N}(T)$ has an exponential decay
\begin{equation}
f^{(n)}_{M,N}(T)\, \,  \sim \,  \,  \,  e^{-\, \kappa(T;M/N) \cdot \, n \cdot \, R}, 
\label{decay}
\end{equation}
where $R^2=\, M^2+N^2$, where $\kappa(T;M/N)$ depends on the ratio $M/N$, and
vanishes, when $T\rightarrow T_c$, as:
\begin{equation}
\kappa(T;M/N)\,\, \sim \,\, \,  |T-T_c|.
\label{kappadiv}
\end{equation}
This exponential decay defines a second length scale which, when
$T\rightarrow T_c$, is infinitely great compared to the lattice length
scale which defines the interaction energy (\ref{nni}). It is worth
noting that the form (\ref{decay}) implies that the angular dependence
of the decay of the correlation functions is the same for $T>T_c$ and $T<T_c$. 

However, as $T\rightarrow T_c$, each term of the form factor expansion
vanishes because $(1-t)^{1/4}$ vanishes, and each $f^{(n)}_{M,N}(T)$
is finite at $T=T_c$.
The singularities of the $f^{(n)}_{M,N}(T)$ are all at
$T=T_c$. These functions satisfy
 Fuchsian linear differential equations~\cite{mccoy2},
and have logarithmic singularities at $T=T_c$ where the highest power
of $\ln |T-T_c|$ is $n$. It therefore requires a detailed infinite summation  
of powers of logarithms to reproduce the behavior (\ref{short}) for
fixed  $M,N$ as $T\rightarrow T_c$. In fact, this type of
demonstration has never been carried out.

\subsection{Differential equations for 
$\langle \sigma_{0,0}\sigma_{N,N}\rangle$}

The correlation functions $\langle \sigma_{0,0}\sigma_{M,N}\rangle$ are
holonomic (D-finite): they satisfy a set of partial linear differential
 equations in the variables $\sinh E^{v,h}/k_BT$. This is exactly the
 holonomic
property of Feynman integrals, shown by 
Kashiwara and Kawai~\cite{kk}. For 
the diagonal case the more specialized result
holds that $\langle \sigma_{0,0}\sigma_{N,N}\rangle$ satisfies a
linear Fuchsian equation. The order of this linear
differential equation is $N+1$, which is
the minimum order needed to accommodate the singular terms at $T=\, T_c$,
$\, \ln ^p|T-T_c|$ with $0\leq p \leq N$ of (\ref{short}), which were
directly obtained from the determinantal representation (\ref{detdn}).  
A few examples are given in~\cite{mccoy6}.

However the  diagonal correlation $\langle \sigma_{0,0}\sigma_{N,N}\rangle$
has the much more remarkable property, discovered by 
Jimbo and Miwa~\cite{jm} in 1980,
 that it satisfies a second order non-linear differential equation 
\begin{eqnarray}
\hspace{-0.7in}&&\Bigl( t \cdot \, (t-1) \cdot \, \frac{d^2\sigma}{dt^2} \Bigr)^2
\,\, =\,\, 
\nonumber  \\
\hspace{-0.7in}&& \, \, \, \,
 N^2 \cdot \, \Bigl( (t-1) \cdot \, \frac{d\sigma}{dt}\, -\sigma \Bigr)^2
\, -4 \, \frac{d\sigma}{dt}\cdot \,
\Bigl((t-1)\frac{d\sigma}{dt}\, -\sigma \, -{{1} \over {4}}   \Bigr) \cdot \,
\Bigl(t\frac{d\sigma}{dt}-\sigma   \Bigr). 
\label{pvi}
\end{eqnarray}
For $T >T_c$ the diagonal correlation is related to  $\sigma$  by
\begin{eqnarray}
\label{sigma-bas}
\sigma(t)\,\,=\,\,\,\, t \cdot \,(t-1) \cdot \,
 {\frac{d}{dt}}\log\langle\sigma_{0,0}\sigma_{N,N}\rangle \, \,\, -{{1} \over {4}},
\end{eqnarray}
with the boundary condition at $t=0$ 
\begin{equation}
\langle\sigma_{0,0}\sigma_{N,N}\rangle\,\, =\,\,\,
 t^{N/2} \cdot \,\frac{(1/2)_N}{N!}\,\, +O(t^{1+N/2}),
\end{equation}
and, for $T <T_c$, it is related to  $\sigma$  by
\begin{eqnarray}
\label{sigma-haut}
\sigma(t)\,\,=\,\,\, t \cdot \,(t-1)  \cdot \,{\frac{d}{dt}}\log\langle
\sigma_{0,0}\sigma_{N,N}\rangle \,\,\, -{{t} \over {4}}, 
\end{eqnarray}
with the boundary condition
\begin{equation}
\langle\sigma_{0,0}\sigma_{N,N}\rangle
\,=\,\,\, (1-t)^{1/4} \cdot \,
\{1\,-\frac{t^{N+1}}{2N+1}\left(\frac{(1/2)_{N+1}}{(N+1)!}\right)^2
\,\, +O(t^{N+2})\}, 
\end{equation}
where  $\, (a)_N=\,a(a+1)\cdots(a+N-1)$ is Pochhammer's symbol
($N \ge 1$, $(a)_0=1$).
These boundary conditions are obtained from the leading terms of
the form factor expansions as $t\,\rightarrow \,0$.
Equation (\ref{pvi}) is an alternative form of the 
Painlev{\'e} VI equation~\cite{ince}, called the $\, \sigma$-form
of Painlev{\'e} VI.

No nonlinear isomonodromic (Garnier~\cite{Garnier,Okamoto}) systems of
partial differential equations have been derived for the general
correlation function $\langle\sigma_{0,0}\sigma_{M,N}\rangle$, even
though they almost certainly exist. Such isomonodromic systems
would yield, by differential algebra elimination, in the isotropic case,
 higher nonlinear order differential equations
with the Painlev\'e property. Some exact results for 
$\langle\sigma_{0,0}\sigma_{N,N-1}\rangle$ are actually given by 
Witte in~\cite{witte}. 

However
$\langle\sigma_{0,0}\sigma_{M,N}\rangle$ does satisfy quadratic 
difference equations~\cite{mccoy40},~\cite{perk} 
\begin{eqnarray}
\hspace{-0.5in}&&\sinh 2E^h/k_BT \cdot \, \{C(M,N)^2\, -C(M,N-1)\, C(M.N+1)\} 
\nonumber\\
\hspace{-0.5in}&&+\sinh 2E^{*v}/k_BT \cdot \, \{C^*(M,N)^2\, -C^*(M-1,N)\, C^*(M+1,N)\}
\,\,=\,\,\, 0, 
\label{difference1}\\
\hspace{-0.5in}&&\sinh 2E^v/k_BT \cdot \, \{C(M,N)^2\, -C(M-1)\, C(M+1,N)\} 
\nonumber\\
\hspace{-0.5in}&&+\sinh 2E^{*h}/k_BT 
\cdot \, \{C^*(M,N)^2\, -C^*(M,N-1)\, C^*(M,N+1)\}
\,\,=\,\,\, 0, 
\label{difference2}
\end{eqnarray}
where 
\begin{equation}
C(M,N)\,\,\,  =\, \, \,\, \langle\sigma_{0,0}\sigma_{M,N}\rangle , 
\end{equation}
and where $C^*(M,N)$ are the correlations on the dual lattice
(\ref{dual}).
At $T=T_c$, where $\sinh 2E^{*.i}/k_BT=\sinh 2E^{i}/k_BT$ for $i=v,h$,
and $C^*(M,N)=\, C(M,N)$, these equations reduce to
the discrete imaginary time Hirota equation~\cite{Hirota}.

\subsection{Susceptibility}

By using the form factor expansions (\ref{ffm}) and (\ref{ffp}) in  
(\ref{sus}), we obtain the expansion for the 
susceptibility as the infinite sum of $n$ ``particle'' contributions
\begin{eqnarray}
\hspace{-0.5in}&&k_BT \cdot \,\chi_+(T)\,\, = \, \, \,
(1-t)^{1/4} \cdot \,t^{-1/4} \cdot \,\sum_{j=0}^{\infty}\,{\chi}^{(2j+1)}(T), 
\quad \quad \,  {\rm for}~T>T_c, 
\label{chip}\\ 
\hspace{-0.5in}&&k_BT\cdot \,\chi_-(T)\,\, =\,\, \, 
(1-t)^{1/4} \cdot \,\sum_{j=1}^{\infty}{\chi}^{(2j)}(T), 
\quad \quad \quad \quad {\rm for}~T<T_c.
\label{chim}
\end{eqnarray}
where explicit forms of the $\chi^{(j)}(T)$, as $j-1$ dimensional 
integrals for the general
anisotropic lattice, are given in~\cite{ongp}. These series are
convergent for $T \neq T_c$.

We may also consider what we call the ``diagonal susceptibility'',
defined as:
\begin{equation}
k_bT\cdot \,\chi_d(t)\,\,\,=\,\, \, \sum_{N=-\infty}^{\infty}\{\langle
\sigma_{0,0}\sigma_{N,N}\rangle\, -{\mathcal M}^2\}.
\end{equation}
By use of the form of the diagonal form factor expansion 
of~\cite{mccoy2}-\cite{mccoy3} we obtain as analogues
of (\ref{chim}) and (\ref{chip})
\begin{eqnarray}
\hspace{-0.5in}&&k_BT \cdot \,\chi_{d+}(T)\,=\,\,\,
(1-t)^{1/4}\cdot \,t^{-1/4}\cdot \,\sum_{j=0}^{\infty}\,{\chi}_d^{(2j+1)}(t),
\quad \quad {\rm for}~ T>\,T_c,
\label{chidp}\\ 
\hspace{-0.5in}&&k_BT \cdot \,\chi_{d-}(T)\,=\,\,\,
(1-t)^{1/4}\cdot \,\sum_{j=1}^{\infty}\,{\chi}_d^{(2j)}(t),
\quad \quad \quad \quad  {\rm for}~T<\,T_c,
\label{chidm}
\end{eqnarray}
where the $\chi_d^{(n)}(t)$ are given in \cite{mccoy4}. These series
are also convergent for $T\neq T_c$.

The series expansions (\ref{chip}) and (\ref{chim}) 
are, what we referred to in the introduction
as, the analogue of a ``Feynman diagram expansion''.

The integrals for $\chi^{(n)}(T)$, and $\chi^{(n)}_d(T)$, 
have been extensively 
studied~\cite{jm1}-\cite{jm10}. It is
quite instructive to contrast them with the form factor integrals 
$f^{(n)}_{M,N}(T)$ from whence they came.

When $T\rightarrow T_c$ each $\chi^{(n)}(T)$ diverges as
$|T-T_c|^{-2}$ with a coefficient which depends on $n$ and rapidly
decreases as $n$ increases. Thus, in terms of the variable
\begin{equation}
\tau \, \,  =\, \,  \, \frac{1}{2}\cdot \, (\sinh^{-1}2E/k_BT\, -\sinh 2E/k_BT),
\label{taudef}
\end{equation}
we have, for the isotropic lattice~\cite{wmtb},~\cite{mccoy10},\cite{ongp} as
$\tau\rightarrow 0$ 
\begin{equation}
k_BT\cdot \,\chi_{\pm}(\tau)
 \,\,\, \,\, \longrightarrow \,\,\,\,  \, \,\,
{\sqrt 2} \cdot \,C_{\pm} \cdot \,|\tau|^{-7/4}.
\label{chidiv}
\end{equation}
The constants 
$C_{-}$ and $C_{+}$ are different, and are given as infinite series
\begin{equation}
C_{-}\,\,=\,\,\, \sum_{n=1}^{\infty}\,C^{(2n)}, \hspace{.5in}
C_{+}\,\,=\,\,\,\sum_{n=0}^{\infty}\,C^{(2n+1)}, 
\label{csum}
\end{equation}
where the $C^{(n)}$ are   
 $\,C^{(n)}\,=\,\, 2^{-n}\,\pi^{n-1}\,D_n$, with \cite{cran}
\begin{equation}
D_n\,\, = \, \,\,\, \frac{4}{n!}\int_0^{\infty}\, \cdots \,
 \int_0^{\infty}\,\frac{du_1}{u_1}\,
\cdots \,
{{du_n} \over {u_n}} \,\cdot \,
{{  \prod_{i<j} \left( {{u_i-u_j} \over {u_i+u_j}}\right)^2}\over 
{(\sum_{j=1}^n(u_j+1/u_j))^2}}. 
\end{equation}
These integrals have been
studied for  $n=\,1,\,\cdots, \, 6$.
The first terms in (\ref{csum}) have been
 analytically evaluated in~\cite{wmtb}:
\begin{equation}
C^{(1)} \,= \, \, 1, \hspace{.7in} C^{(2)} \,= \, \, \frac{1}{12\pi}. 
\end{equation}
The next leading term was analytically 
 evaluated by Tracy~\cite{tracy1} as
\begin{equation}
C^{(3)}\,\,=\,\,\,\, \frac{1}{2\pi^2} \cdot \,\left(\frac{\pi^2}{3}\,
+2\,-3{\sqrt 3}\,{\rm  Cl}_2(\pi/3)\right) \,  \,
= \, \,\, 8.1446\, \cdots\,  \times \, 10^{-4}, 
\end{equation}
where 
\begin{equation}
{\rm Cl}_2(\theta)\,\,=\,\,\,\,\sum_{n=1}^{\infty}\frac{\sin n \theta}{n^2},
\end{equation}
is Clausen's function. The next term read:
\begin{equation}
C^{(4)}\,\,  =\,\,\,\,  
\frac{1}{16\pi^3}\cdot \, \left(\frac{4\pi^2}{9}\, -\frac{1}{6}
\,-\frac{7}{2}\zeta(3)\right)\,\,
=\,\,2.5448 \,  \cdots \, \times \, 10^{-5}.
\end{equation}
Accurate numerical evaluations have been made~\cite{cran} for $C^{(5)}$ and
$C^{(6)}$:
\begin{eqnarray}
\hspace{-0.3in}&&C^{(5)} \,= \, \,  \,\frac{1}{2^5\pi^4} \, \times \, 0.0024846057\, \cdots\, 
\, =\,\,\,  7.9709118\, \cdots
  \, \times \, 10^{-7}, \\
\hspace{-0.3in}&&C^{(6)} \, \,=  \,\, \, \frac{1}{2^6\pi^5}\,\times \,
  0.0004891422\, \cdots\,\, =\,\,\,  2.497501\, \cdots\, \times \,  10^{-8}.
\end{eqnarray}
A curious feature of these results is that the ratio $C_{+}/C_{-}$ is
found to be closely approximated by $12\pi$, 
and the succeeding terms decrease
by approximately three orders of magnitude.
The study of the  constants $C_{-}$ and $C_{+}$ has been 
continued by high precision numerical computations~\cite{ongp}, 
and the most recent evaluation~\cite{cgnp}, in 2011, is to  104 places. 

For the diagonal susceptibility each $\chi^{(n)}_d(t)$ diverges as
$(1-t)^{-1}$. One finds~\cite{mccoy21},~\cite{mccoy4}
\begin{equation}
\hspace{-0.3in}k_BT \cdot \,\chi_{d+}(T)
\,\, \,\, \, \longrightarrow \,\, \,  \, \, \, \, 
\frac{(1-x^2)^{1/4}}{1-x} \cdot \, 
\sum_{n=0}^{\infty}\,C^{(2n+1)}_{d+},
\end{equation}
where  $\, C^{(1)}_{d+}=\,1$, and
\begin{equation}
\hspace{0.1in}C^{(3)}_{d+}\,\,= 
\,\,\,   \frac{1}{3}\,\,\, 
-\frac{5\pi}{9\Gamma^2(5/6)\Gamma^2(2/3)}\,
-\frac{8\pi}{\Gamma^2(1/6)\Gamma^2(1/3)}
 \, \,=\,\, \,  0.016329 \,\, \cdots
\end{equation}
and 
\begin{equation}
\hspace{-0.3in}k_BT\cdot \, \chi_{d-}
\, \, \, \,\,  \longrightarrow \, \,  \, \, \, \, \, 
(1-t)^{-3/4} \cdot \, \sum_{n=1}^{\infty}\, C^{(2n)}_{d-}, 
\end{equation}
where $C^{(2)}_{d-}\,=\,\,\frac{1}{4}$, and 
\begin{equation}
C^{(4)}_{d-} \, 
= \,  \,\, \,
 \frac{1}{8} \cdot \, \left(1\, -\frac{1}{3\pi^2}[64 \, -16 \, I]\right), 
\end{equation}
with:
\begin{equation}
I\,\, =\, \,\,  -  2.2128121 \,\, \cdots
\end{equation}

In contrast to the form factors $f^{(n)}_{M,N}(T)$, whose only singular point
is $T=\, T_c$ where $T_c$ is real, the $\chi^{(n)}(T)$'s have many further
singularities. The first set of these singularities was found, by 
Nickel~\cite{nic1},~\cite{nic2}, to be, for the isotropic case $E^v=E^h=E$, 
located at 
\begin{equation}
\label{location}
\cosh^{2}2E/k_BT\,\,\, -\,\sinh 2E/k_BT \cdot \, (\cos(2\pi j/n)\,+\cos(2\pi l/n)
\,\,\, =\,\,\, \, 0, 
\end{equation}
with ($[x]$ is the integer part of $x$)
$0\,\leq\, j,\,\,\, \,  l \,\leq \, [n/2], \, 
 j=l=\,0$  excluded (for $n$ even $j+l=\,n/2$ is
also excluded).  
Equivalently (\ref{location}) 
reads $\,\,\, \sinh 2E/k_BT_{j,l}\, =\, \, s_{j,l}$
 $\,\, =\, \,1/2 \cdot \, (\cos(2\pi j/n)+\cos(2\pi l/n))\, $ 
$ \,\, \pm i/2 \cdot \, [(4-(\cos(2\pi j/n) \, +\cos(2\pi l/n))^2]^{1/2}$.
These Nickel's singularities  are clearly
 on the unit circle 
$\, |s| \, =\, 1$, or $\, |k| \, =\, 1$. 
Do note that this is no longer the case for the anisotropic 
model. In the anisotropic case 
Nickel's singularities for the anisotropic  $\chi^{(n)}$'s 
become (see (3.22) of~\cite{ongp}): 
\begin{eqnarray}
\label{location2}
\hspace{-0.4in}&&\cosh^{2}2E^v/k_BT \cdot \, \cosh^{2}2E^h/k_BT \,\, \\
\hspace{-0.4in}&&\qquad \quad  -\,(\sinh 2E^v/k_BT \cdot \, \cos(2\pi j/n)\,
+ \sinh 2E^h/k_BT \cdot \,  \cos(2\pi l/n))
\,\,=\,\,\, \, 0, \nonumber 
\end{eqnarray}
with $\, j$,  $\, l\, = \, \, 1, \, 2, \, \cdots \, n$. These (complex) 
algebraic curves (\ref{location2}), 
in the two complex variables
$\, \sinh 2E^h/k_BT$, $\, \sinh 2E^v/k_BT$, are actually 
singular loci\footnote[2]{This 
result on the anisotropic Ising model has been obtained from 
a Landau singularity analysis,
generalizing~\cite{jm5,jm5s} (S. Boukraa, S. Hassani 
and J-M. Maillard, unpublished
results).} for the D-finite system
of PDE's satisfied by the anisotropic $\, \chi^{(n)}$'s.
These algebraic curves accumulates with 
increasing values of $\, n$, in the same way Nickel's singularities
(\ref{location}) accumulate on the unit circle 
$\, |s| \, =\, 1$, in a certain (real) submanifold $\, {\cal S}$  of the 
two complex variables $\, \sinh 2E^h/k_BT$,
$\, \sinh 2E^v/k_BT$ (four real variables). However, this ``singularity
manifold'' $\, {\cal S}$ is not a codimension-one (real) 
submanifold (like the unit circle 
$\, |s| \, =\, 1$ in the $\, s$-complex plane), but 
actually a codimension zero submanifold, 
as can also be seen on various analyses of complex temperature zeros 
(see for instance~\cite{wang}). Note that this  ``singularity
manifold'' becomes very ``slim'' near the (critical) algebraic curve
$\, k \, = \, \, \sinh 2E^h/k_BT \cdot \,  \sinh 2E^v/k_BT \, = \, \, 1$
(see for instance figures 1, 2 and 3 near the real axis in~\cite{wang}).
The relation between the analyticity in the two complex variables 
$\sinh 2E^h/k_BT$ and $\sinh 2E^v/k_BT$, and the single complex
variable $T$ with $E^v$ and $E^h$ real and fixed,
 will be discussed elsewhere.

\vskip .2cm 

Back to the isotropic model, introducing the variable 
\begin{equation}
w\,\,\,=\,\,\, \, 
\frac{1}{2\cdot \, (\sinh 2E/k_BT \,+(\sinh 2E/k_BT)^{-1})}, 
\end{equation}
the singularities (\ref{location})  for $n=\, 3,4,5,6$ are given in table 1,
where we note, when $\, w$ is real, that $\sinh 2E/k_BT$ is real for 
$-1/4\, \leq \, w\, \leq \, 1/4$,
 and is complex with $|\sinh 2E/k_BT|=\, 1$ for $|w| > 1/4$. 
Following~\cite{nic2} we define $\epsilon$, the deviation from 
the singular temperatures
$T^{(n)}_{j,l}$ determined by (\ref{location}), as
$\, s^{-1}\,\,=\,\,\, (1-\epsilon)\cdot \, s^{-1}_{j,l}$, for $~~T<T_c$,
and $\, s\,\,=\,\,\, (1-\epsilon) \cdot \, s_{j,l}$,   for $~~T>T_c$. 
Then, for $T<\, T_c$, 
the singularity in  ${\chi}^{(2n)}(T)$, at $T_{j,l}$, reads 
\begin{equation}
A^{(2n)}_{j,l}\cdot \, \epsilon^{2n^2-3/2}.
\label{singeven}
\end{equation}
and, for $T>T_c, $ 
the singularity in
${\chi}^{(2n+1)}(T)$ reads 
\begin{equation}
A^{(2n+1)}_{j,l}\cdot \, \epsilon^{2n(n+1)-1} \cdot \,\ \frac{\ln \epsilon}{\pi}. 
\label{singodd}
\end{equation}  
The amplitude $A^{(N)}_{j,l}$ is given by~\cite{nic2}
\begin{eqnarray}
A^{(N)}_{j,l}&=& \, \frac{(N i\,\sin\theta_{j,l})^{(N^2-3)/2}}
{(\sin^2\phi^{(j)}\,\cos \phi^{(l)}\,+\sin^2\phi^{(l)}\,\cos\phi^{(j)})^{(N^2-1)/2}}
\nonumber\\
&\times&
\frac{\prod_{m=1}^{N-1}(m!/2^m)}{\pi^{(N-3)/2)}
\, \sqrt{N}\,\Gamma((N^2-1)/2)}, 
\label{amp}
\end{eqnarray}
with 
\begin{equation}
\phi^{(m)} \, \,  =\, \,  \, 2\pi m/N, \quad \quad \quad 
2\cos\theta_{j,l} \,  \, = \, \, \,   \cos\phi^{(j)}+\cos\phi^{(l)}.
\end{equation}

\begin{table}[h!]
\caption{The Nickel singularities (\ref{location}) for the isotropic 
case $E^v=E^h$ of $\chi^{(n)}$ for $n=\,3,4,5,6$}

\begin{center}
\begin{tabular}{|l|l|}\hline
$n$&$w$\\ \hline
3&$-1/2,~1$\\
4&$\pm 1/2$\\
5&$-1,~\frac{-1\pm \sqrt{5}}{4},~\frac{3\pm \sqrt{5}}{2}$\\
6&$\pm 1,~\pm 1/3$\\ \hline
\end{tabular}
\end{center}
\end{table}

The Fuchsian linear differential equations~\cite{jm1}-\cite{jm10}
 which the (isotropic)
$\chi^{(n)}(T)$  satisfy, have singularities which accumulate
 on the unit circle, but, also, inside, 
and outside, this unit circle $|\sinh 2E/k_BT|=\, 1$. 
However, the series expansion of these integrals $\chi^{(n)}(T)$
are actually analytic inside the unit circle ($|\sinh 2E/k_BT|<\, 1$).
The singularities of a linear ODE and the singularities of 
a particular series-solution of an ODE do not coincide. 

The singularities of the Fuchsian ODE may be obtained, from the 
integrand of the integrals, by the same ``Landau''
analysis~\cite{jm5} used to obtain singularities of Feynman diagrams. 

For the isotropic $\chi^{(3)}$, in addition to the unit circle 
singularities (\ref{location}), 
there are additional singularities at
\begin{equation}
w\,\,= \,\, \,\frac{-3 \pm \,i\,{\sqrt 7}}{8}, 
\quad \quad {\rm where} \quad \quad \, \,  
s,\,\,s^{-1}\,\, =\,\, \frac{-1\,\pm \,i\,{\sqrt 7}}{2},
\end{equation}
and they actually correspond 
to complex multiplication of elliptic curves,
Heegner numbers, and complex fixed points
 of the Landen transformation~\cite{jm5s}.

The singularities of $\chi^{(n)}(w)$ are to be contrasted with 
the diagonal susceptibility $\chi^{(n)}_d(t)$ which only have~\cite{jm5}
singularities on the unit circle $|t|=\, 1$.

For $T<T_c$ the singularities of $\chi_d^{(2n)}(t)$ are at 
$t_0= \, e^{2\pi ij/n}$, and are of the form:
\begin{equation}
A^{(2n)}_{d;j} \cdot \, \epsilon^{2n^2-1} \cdot \,\ln \epsilon.
\end{equation} 
For $T>T_c$ the singularities in $\chi_ d^{(2n+1)}(t)$ are at
$t_0\,= \, e^{2\pi j/(n+1/2)}$, and are of the form
\begin{equation}
A^{(2n+1)}_{d;j}\cdot \, \epsilon^{(n+1)^2-1/2},
\end{equation}
where the amplitudes $A^{(N)}_{d;j}$ have yet to be determined.

The linear differential operator for $\chi^{(3)}$ rightdivides 
the one for $\chi^{(5)}$ (in a direct sum structure~\cite{jm6}).
Consequently, all the singularities of the linear 
differential operator for $\chi^{(3)}$ are also singularities
of the  operator for $\chi^{(5)}$.
It was, however,  seen~\cite{jm6}, by
means of a Fast Fourier Transform analysis of the
series expansion of the integral
for $\chi^{(5)}$, as a power series in $t$, that some $\chi^{(3)}$
singularities are not present in $\chi^{(5)}$. This is to be
contrasted with a very recent result~\cite{hass} for 
the diagonal susceptibility $\chi^{(5)}_d$ which has found that 
the singularities of $\chi^{(3)}_d$ are present in $\chi^{(5)}_d$.

\subsection{Natural boundary for the isotropic model}

It is striking that the number of singularities (\ref{location}) 
in $\chi^{(n)}(T)$ increases with $n$, and becomes dense in the limit
 $n \, \rightarrow  \, \infty$. This 
feature led Nickel~\cite{nic1} to the conclusion that, unless cancellations
occur, there will be a unit circle natural boundary 
in the susceptibility $\chi(T)$
in the complex $s$ plane. The fact that ref.\cite{jm6} demonstrates
that some singularities  of the $\chi^{(3)}$ series
are not singularities of the $\chi^{(5)}$
series, supports the no cancellation assumption.   
The existence of a natural boundary, in the complex
temperature plane, is not contemplated in the scaling theory of
critical phenomena.  
The literature on natural boundary is quite narrow~\cite{Breuer}, 
as well as the
methods and tools to study the neighborhood of a natural boundary.
Curious situations may occur, like, for instance, the ``radial
porosity'' one encounters with Borel's examples of monogenic 
functions (see section 10.5 page 21 of~\cite{Hille}).

There is much that is, as yet, not understood about the properties of
this Ising natural boundary. In the vicinity of 
any point on $|s|=1$, such that $s\neq \pm1$,
 the local spacing of the singularities, in $\chi^{(n)}$, is of the order 
$n^{-2}$. However, for $s=\pm 1$ the local spacing is only $n^{-1}$, 
and the dependence of the amplitude (\ref{amp}), for
large $N$, is different in these two cases.

An initial analysis was made in~\cite{ongp}, for $s \, \rightarrow  \, 1$,
based on keeping only the singularities closest to $\tau= \, 0$. 
This analysis
yielded an essential singularity of the form $e^{-C/\tau^2}$ where $C$ is a
constant. However, there is more to a natural boundary than just an
essential singularity, and further analysis will be required to fully 
assess the properties following from the singularities in $ \, \chi^{(n)}$.
In particular we note that, in general, the limiting value at a point on
a natural boundary depends on the direction of the approach, and, as
suggested in~\cite{ongp},
an asymptotic expansion about $|s|=1$, $ \, s\neq\,  \pm1$, may not exist.
The existence of a unit circle natural boundary in the isotropic
square Ising model, seriously questions most of the scaling arguments 
taken for granted on this model. For example it is desirable to
explore the relation between the equality  
$\, \gamma_{+}\, = \,  \, \gamma_{-}  \,$
for the susceptibility exponent $\, \gamma$, above and below $T_c$, 
and the unit circle natural boundary which 
disconnects\footnote[3]{As far as 
analysis of several complex variables~\cite{Fuks}
is concerned,
 the situation is even worse
for the anisotropic Ising model, because of the codimension zero
manifold $\, {\cal S}$ of section 3.6.} the 
inside and the outside of the unit circle, 
$\, |k| < 1 $ and $\, |k| > 1 $.

 \subsection{Series expansions}

There is a second way of studying the susceptibility which is distinct from
(and in a way complementary to) the use of the form factor
expansion. By use of either (in principle) the determinant
representation or (in practice) by the
more sophisticated approach of using the nonlinear difference equations
(\ref{difference1}), (\ref{difference2}) with the explicit result for 
$\langle \sigma_{0,0}\sigma_{N,N}\rangle$ of (\ref{diagexplicit}), and
for $\langle \sigma_{0,0}\sigma_{N,N-1}\rangle$ of~\cite{witte},
very long expansions about $T=\, 0$, and
$T=\, \infty$, can be obtained.
For the isotropic lattice $E^v=E^h=E$ this has been done 
in~\cite{ongp} and~\cite{cgnp}, and series expansions with may hundreds
of terms have been obtained,
not only on the square, but also on the triangular and
honeycomb lattices. 

These long expansions have been analyzed in~\cite{ongp} and~\cite{cgnp}.
This analysis leads to the following
expansion of the susceptibility as $T\,\rightarrow \,T_c$ which
generalizes the leading diverging term given in (\ref{chidiv}) 
 \begin{equation}
k_BT \cdot \, \chi_{\pm}\,\, =\, \,\, \,  \, 
 C_{\pm} \cdot \,| \tau|^{-7/4} \cdot \,F_{\pm}(\tau)\,\,\, + \, B(\tau), 
\label{series}
\end{equation}
where
\begin{equation}
F_{\pm}(\tau)\,\, =\,\,\,  1\,\,\, 
+\sum_{n=1}^{\infty}\,f_{n\pm}\cdot \,\tau^n, 
\label{fpm}
\end{equation}
and
\begin{equation}
B(\tau)\,\, =\,\,\,\,\,  \sum_{q=0}^{\infty}\sum_{p=0}^{[{\sqrt q}]}\,
b^{(p,q)}\cdot \,(\ln \tau^2)^p \cdot \,\tau^q, 
\label{chishort}
\end{equation}
with $\tau$ taken to be real, and where the $b^{(p,q)}$'s are the 
same for $T<T_c$ and $T>T_c$,
and are functions depending on the lattice. Unlike the expansions
(\ref{chip}), and (\ref{chim}), which converge we will see below that
(\ref{chishort}) is expected to only be asymptotic. The function
$F_{\pm}(\tau)$ is referred to, in~\cite{ongp} and~\cite{cgnp}, as a
``scaling function''. 

It is greatly instructive to compare the result (\ref{series}) of~\cite{ongp} 
and~\cite{cgnp} with the behavior of the expansions
(\ref{chip}) and (\ref{chim}), derived from the form factor expansion.

The term $B(\tau)$ is of the same form  as the logarithmic terms (\ref{short})
already seen in the determinantal form of the correlations. Such terms
are present term by term in (\ref{sus}), and are referred to, 
in~\cite{ongp}, as ``short distance'' contributions. However, in the expansion
(\ref{chip}) and (\ref{chim}), the $\chi^{(n)}(T)$ have singularities,
at $T\rightarrow T_c$,  with powers of $\ln|T-T_c|$. Thus, term by
term in (\ref{chip}) and (\ref{chim}), the coefficient $(1-t)^{1/4}$ is
not cancelled out.  Consequently there must be an infinite number of
detailed cancellations to obtain the $B(\tau)$ of (\ref{chishort}) from 
(\ref{chip}) and (\ref{chim}).

The most interesting question  concerns the convergence 
of the infinite series in
(\ref{fpm}) and (\ref{chishort}). This cannot, of course, be answered
by a finite series, no matter how long and, in fact, the numerical
results for $F_{\pm}(\tau)$, and $b^{(p,q)}$ in~\cite{cgnp}, 
do not show  divergent
behavior with the number of terms which have been computed. However, the
dense set of singularities in the $\chi^{(n)}(T)$, found 
by Nickel~\cite{nic1},~\cite{nic2} 
in analyzing the form factor integrals, must
have a significant influence on this expansion. The  influence
of Nickel's singularities has been analyzed in~\cite{ongp} with the
conclusion that there must be  asymptotic behavior in, at least, some
of the series in (\ref{fpm}) and (\ref{chishort}). 
It is argued in~\cite{ongp}, from the results of (6.12),
that the coefficients $b^{(p,0)}$ form an asymptotic sequence for
sufficiently large $p$. The behavior of $b^{(p,q)}$, for $p$
fixed and $q \rightarrow \infty$, and $f_{n\pm}$ as 
$n \, \rightarrow \, \infty$, remains to be carried out. The further effects
  caused by the Landau singularities~\cite{jm5} also remain to be studied.

\subsection{The scaling (field theory) limit at $H=0$}

One of the most important properties of the Ising model at $H=0$, 
as $T\, \rightarrow \,T_c$, is
that  it defines a Euclidean quantum field
theory and gives a very concrete example of the concepts of mass and
wave function renormalization.  

We concretely illustrate this for the two point correlation 
$\langle\sigma_{0,0}\sigma_{M,N}\rangle$.
For mass renormalization we define
\begin{equation}
r\,\,\, =\,\,\,\,  \kappa(T;M/N) \cdot \, R,
\label{scaledr}
\end{equation}
where $R^2=\,M^2 +N^2$, and where $\kappa(T;M/N)$ 
is the inverse correlation length introduced in
subsection (3.4). We define what we call the scaling limit, for $T$
real, as
\begin{equation}
\hspace{-0.5in}T\,\longrightarrow \,T_c, \qquad \quad R\,\longrightarrow \,\infty,
\label{slimit}
\end{equation}
with $r$ fixed, and we recall from (\ref{kappadiv}) that $\kappa$
vanishes as
$T\,\rightarrow \,T_c$ 
\begin{equation}
\kappa(T;M/N) \,\,\, \sim \,\, \, \,A_{\kappa}\cdot \, (1-t).
\label{kdiv}
\end{equation}

For wave function renormalization we divide the correlation function
by the factor $(1-t)^{1/4}$ which vanishes at $T=\,T_c$. When $T<T_c$
this factor is the square of the spontaneous magnetization. The
interpretation is that the spins, which on that lattice have the values
$\sigma=\pm1$, are regarded as having a ``natural size'' of
$\mathcal{M}$ in the scaling limit. Similarly, for $T>T_c$, the value of
the disorder parameter $M_{+}$ is interpreted as the natural 
size for $\sigma$.   

We then define for $T\,\rightarrow\, T_c\pm$ 
\begin{equation}
\hspace{-0.3in} G_{\pm}(r)\,\,=\,\,\, \,
\lim_{\rm scaling}(1-t)^{-1/4} \cdot \,\langle\sigma_{0,0}\sigma_{M,N}\rangle.
\label{green}
\end{equation}
By use of the form factor expansion we see that this limit exists. In
the isotropic case this function is rotationally invariant.
In the anisotropic case it becomes rotationally invariant 
if one uses the length variable $\, r$ (fixed)
\begin{equation}
\left[\left(\frac{\sinh 2 E^h/k_BT_c}{\sinh 2 E^v/k_BT_c}\right)^{1/2}\cdot \,M^2\,
+\left(\frac{\sinh 2 E^v/k_BT_c}{\sinh 2  E^h/k_BT_c}\right)^{1/2}
\cdot \, N^2\right]^{1/2}(1-t)\,\, =\,\,\, r.
\label{rdef}
\end{equation}

This function $G_{\pm}(r)$ is expressed in terms of a Painlev{\'e} equation 
of the third kind 
\begin{equation}
\frac{d^2\eta}{d\theta^2}\,\,\, =\,\,\,\,
\frac{1}{\eta}\left(\frac{d\eta}{d\theta}\right)^2 \,
\,-\frac{1}{\theta}\frac{d\eta}{d\theta}\,\,+\eta^3 \, -\eta^{-1},
\label{pviiiform}
\end{equation}
as
\begin{equation}
\hspace{-0.3in} G_{\pm}(r)\,\,=\,\,\,\, \, 
\frac{1\, \mp \,\eta(r/2)}{2\,\eta(r/2)^{1/2}}\cdot \, 
\exp\frac{1}{4}\int_{r/2}^{\infty}\,d\theta \cdot \, \theta \cdot \, 
\frac{(1-\eta^2)^2\,-(\eta')^2}{\eta^2}.
\label{pviiiformG}
\end{equation}
The second order equation (\ref{pviiiform}) admits a one parameter
family of solutions which decay exponentially, when 
$\theta \rightarrow  \infty$, as
\begin{equation}
\eta(\theta)\, \, \sim \,\,\, \, 
1\,\, -\frac{2\lambda}{\pi}\cdot \,K_0(2\theta)
\label{boundarycond}
\end{equation}
where $K_0(2\theta)$ is the modified Bessel function.
The specific value $\lambda=1$ corresponds to the Ising model.
The  result was first announced in~\cite{mccoy10} and~\cite{mccoy20}. Two
different proofs were originally given. The first, in~\cite{wmtb}, is based on 
Myers' work~\cite{myers}, on the scattering of
electromagnetic radiation from a strip, and the second~\cite{mtw} is
based on a direct manipulation of the exponential representation
in the scaling limit.

An alternative derivation of the scaled two-point function
was subsequently obtained~\cite{jm}  by directly taking
the scaling limit of the Painlev{\'e} VI equation (\ref{pvi}). This
scaling leads to  representation in terms of a Painlev{\'e} V function. 
The equivalence of this representation with the original  Painlev{\'e}
III form was shown in~\cite{mp}.

The scaling limit as defined by (\ref{scaledr}) and
(\ref{slimit}), which is used to define the scaled Green's function
(\ref{green}), is defined from the massive regime where the
correlation on the lattice decays exponentially. It remains to connect
this regime with the algebraic decay of the lattice correlations,
at $T=\, T_c$, given by (\ref{corrtc}). We do this by extending the strict
limiting definition (\ref{green}) with the less precise statement that
\begin{equation}
\langle \sigma_{0,0}\sigma_{M,N}\rangle \,\, \sim \, \,\, 
(1-t)^{1/4}\cdot \, G_{\pm}(\kappa \, R), 
\label{moreg}
\end{equation}
and examine (\ref{moreg}) as $T \rightarrow T_c$, i.e. $t\rightarrow 1$ and 
$\kappa \rightarrow \infty$. Thus, we see that if, as $r\rightarrow 0$, we
have
\begin{equation}
G_{\pm}(r) \,\,  \,  \,   
 \longrightarrow \, \,\, \,\, \, \,   \,     \,   A_G/r^{1/4}, 
\label{gform}
\end{equation}
then (\ref{moreg}) reduces to
\begin{equation}
\langle \sigma_{0,0}\sigma_{M,N}\rangle \,\,  \,   \,  
 \longrightarrow  \,\, \,\,  \,  \,   \,   
 A_G \cdot \,  \left(\frac{A_{\kappa}}{R}\right)^{1/4}.
 \end{equation}
The exponent $1/4$ in (\ref{gform}) is shown in~\cite{mtw} to follow, from 
the local behavior of  the Painlev{\'e} III equation at
$\theta\rightarrow 0$, if the exponentially decaying boundary condition
(\ref{boundarycond}), at infinity, holds with $\lambda=1$. 

The constant
$A_G$ does not follow from a local property of the Painlev{\'e} function.
and the limit of $G_{\pm}(r)$, as $r\rightarrow 0$, will agree with
$\langle \sigma_{0,0}\sigma_{M,N}\rangle$, as $R\rightarrow \infty$, if
in addition to the functional form (\ref{gform}), it can be shown that
\begin{equation}
A_G \cdot A^{1/4}_{\kappa}\,\,  \,  =\,\, \, \,  \,    A_c, 
\label{equality}
\end{equation}
with $A_c$ given by (\ref{ac}). This crucial identity was first proven
by Tracy~\cite{tracy2} in 1991, by use of the scaling limit of the
exponential forms of the correlation (\ref{expm}) and (\ref{expp}).
It is perhaps worth pointing out that a derivation of the constant 
$A_G$ of (\ref{gform}) has never been directly obtained from the Painlev{\'e}
III form (\ref{pviiiform}), (\ref{pviiiformG}).

\section{The Ising model for $H \neq 0$}

The properties of the Ising model at $H=\, 0$, presented in the previous
section, are all found by exact computations which start with the
definition (\ref{nni}) of the nearest neighbor Ising model  
and are mathematically 
rigorous. However when we extend our considerations to $H \neq 0$, 
this is not the case, and, with only a few exceptions, the results
require some arguments which, while often extremely plausible, in fact
include assumptions which remain to be proven. Nevertheless the work
of the last 50 years has produced remarkable results which give a
compelling scenario of the behaviour of the Ising model for $H\neq 0$. 
In this section will here discuss the following major contributions:

\vskip .1cm 
1. Perturbation for small $H$ for the two-point
 function~\cite{mccoy52} and 
two-body 
\vskip .1cm 
$\qquad$ scattering for $T>\, T_c$ ~\cite{zz},

2. Conformal field theory~\cite{bpz},
 
3. Integrable perturbations of conformal field 
theory~\cite{zam},~\cite{zam2},

4. The connection of integrable perturbations 
with Rogers-Ramanujan 
\vskip .1cm 
$\qquad$ identities~\cite{mccoy50}-\cite{wp},

5. Scaling theory with irrelevant operators and nonlinear scaling
\vskip .1cm 
$\qquad$ fields~\cite{ongp},~\cite{cgnp},~\cite{af},~\cite{caselle},

6. Non-integrable perturbations of conformal field theory~\cite{fz}.

\subsection{Perturbation for $H\sim 0$}

For $T<T_c$ the two point function has been 
studied perturbatively~\cite{mccoy52}
 for small $H$. It was found that if in the limit $H \rightarrow \, 0$ and
 $T\rightarrow T_{c-}$ from the low temperature side 
with the ``scaled magnetic field $h$ fixed
\begin{equation}
h\,\,  =\, \,\,\,  \lim_{\rm scaling}\frac{H}{|T-T_c|^{15/8}}, 
\label{hscalingdef}
\end{equation}
then 
the connected part of the two-point function, for $h \sim  \, 0$ 
and large $r$, is
\begin{equation}
G^c(r;h) \, \,  \,   \sim \,  \,  \,  \, \,\,  
\pi^{1/2}  \cdot  \, r^{-1/2}  \cdot  \, e^{-2r} \cdot  \,
 \sum_j a_j(h)  \cdot  \, e^{-r\kappa_j(h)}
\label{gsmallh}
\end{equation}
with
\begin{equation}
\kappa_j(h) \, = \, \, \,   h^{2/3} \cdot  \, \lambda_j 
\quad \quad \quad {\rm and} \quad \quad \quad \, \, 
a_j(h)\,\, =\,\,  h \cdot \, a, 
\end{equation}
where $a$ is a constant, and the $\lambda_j$'s are solutions of
\begin{equation}
J_{1/3}(\lambda^{3/2}/3) \,\, +J_{-1/3}(\lambda^{2/3}/3)
\,\,\,\, =\,\,\,\,\, 0, 
\end{equation}
where $J_n(z)$ is the Bessel function of order $n$.
  
This perturbation is, in  fact, a ``singular'' perturbation which is
subject to some interpretation. In particular one is not able to
distinguish between $r^{-1/2} \,e^{-\kappa \;r}$ and $K_0(\kappa\; r)$, where
$K_0(z)$ is the modified Bessel function. If we make this replacement in 
(\ref{gsmallh}), then the Fourier transform consists of a set of poles,
and this result can be interpreted as an example of confinement of  
two ``domain wall excitations''  which interact by means of an infinitely 
weak linear confining potential, to produce a spectrum of ``mesons''. 
For $T>T_c$ the only effect on the leading behavior of the two-point
function, for $h\sim 0$, is to shift the correlations length by a 
term proportional to $h^2$. 
From these computations a scenario is conjectured in~\cite{mccoy52} 
that, as we go from
$T<T_c$ to $T>\, T_c$ in a path in the $(H,T)$ plane in the scaling limit, 
the Fourier transform of the two-point function will contain 
poles. These poles can be interpreted as quasi particles and  the number
of these poles will go, in a smooth fashion, from one at $T>T_c$, $\, H= \,0$,
to the infinite number of poles for $T<T_c$ and $h\rightarrow\,  0$ given by
(\ref{gsmallh}).

Much more recently~\cite{zz} the two-body scattering has 
been studied in the same small $h$ limit for $T>T_c$. 
One of the very significant results of this study is that, at large
energies, the scattering is predominantly inelastic. 
 
\subsection{Conformal field theory}

Conformal field theory is an entirely new approach to critical
phenomana invented by Belavin, Polyakov and Zamolodchikov~\cite{bpz} in 1984. 
In this approach there is no lattice such as (\ref{nni}) and the
theory is defined directly in the continuum. In particular, these 
continuum theories make no contact with
the short distance terms, and the natural boundary, which were discussed
in section 3.   

It is not our purpose here to present a survey of conformal field theory,
which is well presented in the original paper~\cite{bpz} 
 and, subsequently, in many places~\cite{difr},~\cite{mussardo}.
Instead, we will restrict our attention to a discussion of 
two integrable perturbations relevant to the Ising model, the $M(3,4)$
and the $M(2,5)$ minimal models. 

The minimal model $M(3,4)$ describes the critical point of the 
Ising model at $T=T_c$ and $H=0$. It has two relevant operators for
the energy and magnetization. The conformal dimension of energy is
$1/2$, which means the two-point function for energy is $r^{-2}$, and
for magnetization is $1/16$, which means that the two-point spin-spin
correlation is $r^{-1/4}$. These results agree which the exact results at
$T=T_c$ of the Ising  model (\ref{nni}).

There is, however, a second conformal field theory which is relevant
to the Ising model, namely the $M(2,5)$ model which is related to the
Lee-Yang edge. 

In 1952 Lee and Yang \cite{leeyang} proved what is one of the 
few results exactly known for the Ising model in a magnetic field,
namely that, for real interaction energies, the zeros of the partition
function of an isotropic Ising model on a finite lattice, all lie on the
circle $z=\, 1$, where 
\begin{equation}
z \, \,  =\, \,  \, e^{-2H/k_BT}.
\end{equation}
For $T>\, T_c$ these zeros
are all bounded away from the unit circle $z=\, 1$, and they pinch 
the real $z$-axis when $T \rightarrow \, T_c$. For $T<\, T_c$ they fill up
the unit circle $|z|=1$. For $T>T_c$ the endpoint of the arc of zeros is called
the Lee-Yang edge. 

The confining of partition function zeros to an
arc in the complex $z$ plane for real temperatures is to be contrasted 
with the zeros of the partition function in
the complex plane $u=e^{-4E/k_BT}$ for real $H$, where
 computer studies~\cite{shrock},
 on systems of size\footnote[4]{We have also 
performed, with I. Jensen, calculations of 
zeros of partition function of the square Ising model in a 
magnetic field up to size $\, 20 \times 30$.} up to $16\times 16$ 
 show that the zeroes for $H=0$ are located on curves only for very special
boundary conditions~\cite{brascamp}, and for $H\neq 0$ there are regions
of the complex $u$-plane where, even for these special boundary conditions, the
zeros do not lie on curves.

The identification of the Lee-Yang edge as a critical point, with a
continuum field theory interpretation, was first made by Fisher~\cite{fisher},
 and the identification of this field theory with the $M(2,5)$ 
minimal model was first made by Cardy~\cite{ly1}. There is only one
relevant operator and the dimension is $-1/5$, which means that the 
two-point function of this operator is $\langle \phi_0 \phi_r\rangle$ is $r^{4/5}$.
Because there is only one relevant operator, both the energy operator
$\sigma_{j,k}\sigma_{j,k+1}$ and the spin operator $\sigma_{j,k}$
  must have the same two-point function as the operator $\phi$ of the
  $M(2,5)$ model.

The value of the magnetic field $\,H\,=\,\, i\,H_{LY}$,
 at the Lee-Yang edge, vanishes
at $T\rightarrow T_c$ from above,  as:
\begin{equation}
H_{LY}\, \, \, = \, \, \, \, C_{LY} \cdot \,  (T-T_c)^{15/8}.
\end{equation}
Cardy has determined~\cite{ly1} that the density of zeros
$\rho(Im (H))$ diverges, when $\,Im(H) \, \rightarrow \, H_{LY}$, as:
\begin{equation}  
\rho(Im(H)) \, \, \sim  \, \, \,(Im(H) -H_{LY})^{-1/6}.
\end{equation}

\subsection{Integrable perturbations of conformal field theory} 

A perturbation of a conformal field theory is called integrable if it
preserves an infinite number of the constants of the motion of the
conformal field theory which is being perturbed. This concept was introduced
by Zamolodchikov~\cite{zam},~\cite{zam2} at a conference in 1988, where 
he discovered that the perturbation of the $M(3,4)$ model, by both the
energy and separately by the spin operator, have an infinite number of
conservation laws. The $M(2,5)$ model was shown~\cite{ly2}
 to have an integrable
perturbation by the operator $\phi$ in 1990.

\subsubsection{The case of $M(3,4)$}

The perturbation of the minimal model $M(3,4)$ by the energy operator
$\epsilon= \, \sigma_{j,k}\sigma_{j,k+1}$ is, in fact, nothing more 
than saying that the original Ising model
with nearest neighbor interactions (\ref{nni}) is integrable, which is
manifestly true and completely not surprising. However, the discovery
in~\cite{zam}, and~\cite{zam2}, that the perturbation by the spin operator
$\sigma_{j,k}$, which corresponds to the Ising model, at $T=\, T_c$ in a
non zero magnetic field, is integrable came as a big surprise, 
because the lattice interaction (\ref{nni}), at the critical 
temperature with $H\,\, \neq 0$, is
not integrable. Fortunately, this mystery was resolved in 1992
when an integrable lattice model was found~\cite{wns} 
which does have 
the critical behaviour of the magnetization of $H^{1/15}$, 
found in~\cite{zam} and~\cite{zam2}, for the perturbed conformal field
theory. This is the behavior of the Ising magnetization obtained,
decades ago, by simple scaling arguments for critical exponents. 

The truly remarkable property of both, the perturbed conformal field
theory and the lattice realization, is that they have an excitation spectrum
with eight quasi-particles, and that the masses of these particles are
proportional to the components of the Perron-Frobenius eigenvector of
the Cartan matrix of the $E_8$ Lie algebra~\cite{zam},~\cite{zam2}.
This is completely in accordance with the scenario proposed from the
perturbative computation of~\cite{mccoy52}.

\subsubsection{The case of $M(2,5)$}

The integrable perturbation of the $M(2,5)$ model was investigated
 in~\cite{ly2}. In this case there is only  a single quasi particle excitation.
A lattice realization is regime I of the hard hexagon model~\cite{baxhh}.

Because there is only one relevant operator the
identification of this perturbation with the Lee-Yang edge, for real
$T\neq T_c$ and complex magnetic field, should be equivalent to the
corresponding edge in the complex energy (temperature) plan for
real magnetic field $H\neq 0$. However, the complex energy
partition zeros do not lie on curves the way the complex magnetic
field zeros do. Thus, for this identification to hold, further
restrictions on these complex energy zeros must hold in the vicinity
of the edge. Consequently, the precise relation the energy edge has
with the perturbed $M(2,5)$ model remains an open question.

\subsection{Rogers-Ramanujan identities}

All conformal field theories possess a Bose-Fermi 
duality~\cite{mccoy53},~\cite{mccoy54}. The Bose
form gives characters in the Roccha-Caridi form~\cite{rc} by eliminating 
singular vectors from a Bosonic Fock space. For the minimal models
$M(p,p')$, these characters are
\begin{equation}
\chi^{(p,p')}_{r,s}\, = \, \, \,
 \frac{1}{(q)_{\infty}} \cdot \, \sum_{j=-\infty}^{\infty} 
(q^{j(pp'j+rp'-sp)}\, -q^{(p'j+s)(pj+r)}),
\end{equation}
where 
\begin{equation}
(q)_m \,\, =\,\,\, \,\,  \prod_{k=1}^m(1-q^k), 
\end{equation}
for $1\leq r\leq p-1$ and $1\leq s\leq p'-1$.
This is a ``field'' representation of the theory,
 and can be thought of as being
natural to characterize ``short distance'' (ultraviolet) properties. An
equivalent representation is given in terms of exclusion statistics
applied to a set of Fermionic quasi-particles. This gives rise to a
Fermionic form of the characters. These Fermionic forms give a
particle characterization of the spectrum, and can be thought of
as the natural characterization of ``long distance'' 
(infrared) properties.
The relation between the Bose and
Fermi forms is a generalization of the famous Rogers-Ramanujan
identities~\cite{rogers},~\cite{rrh}.

The illustration of this duality for the Ising model, at $T=T_c$, $\, H=0$, as
the $M(3,4)$ minimal model is very instructive.
 
From (2.8) of~\cite{mccoy50} the Fermionic representation of character
$\chi^{(3,4)}_{1,2}$, for the spin operator
of the minimal model $M(3,4)$ which characterizes the Ising conformal
field theory, is
\begin{eqnarray}
\hspace{-0.3in} \chi^{(3,4)}_{1,2}&=&\sum_{m=0\atop m~{\rm odd}}^{\infty}
\, \frac{q^{m(m-1)/2}}{(q)_m}
\label{spinp}\\
\hspace{-0.3in}&=&\sum_{m=0\atop {\rm even}}^{\infty} \, \frac{q^{m(m-1)/2}}{(q)_m}.
\label{spinm}
\end{eqnarray}
Similar formulas hold for the identity
 character $\chi^{(3,4)}_{1,1}$, and the energy character $\chi^{(3,4)}_{2,1}$.
These Fermionic forms match the particle excitations seen in the form
factor expansion of the form factors for $H=\, 0$. There is one type 
of excitation, 
and the index $m$ represents the contribution of a $m$-particle
state. The sum over odd (even) $m$ in (\ref{spinp}) (respectively
 (\ref{spinm})), corresponds to the odd (even) number of
quasi-particles in the form factor expansion of the two-point function
for $T>T_c$ ($T<T_c$). The equality of (\ref{spinp}), and (\ref{spinm}),
 represents the fact that, at $T=\, T_c$, these representations of
 the spectrum for $T\neq T_c$ must give the same result. In the
 language of perturbed quantum field theory this characterization of the
 spectrum corresponds to perturbing the $M(3,4)$ minimal model by the
 energy operator.
 
There is, however, another Fermionic representation of these same characters
which was first conjectured in (2.18) of~\cite{mccoy51}
 (and proven in~\cite{wp}) 
\begin{equation}
\chi^{(3,4)}_{1,1}\,\, =\,\, \,\, 
\sum_{m_1=0}^{\infty}\, \cdots \, \sum_{m_8=0}^{\infty}
\frac{q^{{\bf m}C^{-1}_{E_8}{\bf m}^t}}{(q)_{m_1}\cdots\, (q)_{m_8}}, 
\end{equation}
where ${\bf m}=\, (m_1,\cdots m_8)$, and where $C_{E_8}$ is the Cartan matrix of
  the Lie algebra $E_8$.  Similar identities hold \cite{mccoy53} for 
$\chi^{(3,4)}_{1,1}+\chi^{(3,4)}_{1,2}$ and 
$\chi^{(3,4)}_{1,1}+\chi^{(3,4)}_{1,2}+\chi^{(3,4)}_{1,4}$. 
This representation has eight different types of excitations, which have no
restriction on the parity of the number of allowed 
  excitations. This is in exact correspondence with the eight particles,
  found by Zamolodchikov~\cite{zam} by perturbing the $M(3,4)$
  conformal field theory by the spin operator. 

For $M(2,5)$ the fermionic forms of the characters have only one 
quasi-particle, and read
\begin{equation}
\chi^{(2,5)}_{1,1}\,\, =\, \,\,\,\, \sum_{n=0}^{\infty}\frac{q^{n^2+n}}{(q)_n},
\hspace{.5in}\chi^{(2,5)}_{1,2}\,\, =\,\, \,\, \,
 \sum_{n=0}^{\infty}\frac{q^{n^2}}{(q)_n}, 
\end{equation} 
which are, in fact, the original identities of Rogers~\cite{rogers} 
and Ramanujan~\cite{rrh}. 

 \subsection{Scaling theory}

A completely different approach to the Ising model, with $H\neq 0$,
is the scaling theory of the renormalization group which was developed
before the discovery of conformal and perturbed quantum field theory.
This is presented  in detail
the classic paper of Aharony and Fisher~\cite{af}, and tested
extensively in~\cite{ongp} and~\cite{cgnp}.  

In~\cite{af} 
the conjecture is presented, for $T$ and $H$ real, that
\begin{equation}
F\,\,=\,\,\,\, -g_t^2\cdot \,\ln g_t^2\cdot \,{\tilde Y}\,\,
 +g_t^2\cdot \,Y_{\pm}(g_h/|g_t|^{15/8})\,\,\, +A_0(t), 
\label{af1}
\end{equation}
where the function $A_0(t)$ is analytic at $\tau=\, 0$, and where
$Y_{\pm}(z)$ refer to $T$ above and below $T_c$, and
\begin{eqnarray}
&&g_t\,\,=\,\,\,\,   \sum_{n=0}\, a_{2n}(\tau)\cdot \, 
H^{2n},\qquad \quad \quad \quad  a_0(0)=\,\, 0,  \\ 
&&g_h\,\,=\,\,\,\,   \sum_{n=0}\, b_{2n+1}(\tau)\cdot \, H^{2n+1},
\label{nonlinear}
\end{eqnarray}
with
\begin{eqnarray}
&&a_{2n}(\tau)\,\,=\,\, \,\,   \sum_{j=0}^{\infty}\,a_{2n,j}\cdot \,\tau^{j}, 
\qquad \quad \quad {\rm with}\,\, \quad \quad \quad a_{0,0}=\, 0, 
 \nonumber\\
&&b_{2n+1}(\tau)\,=\,\,\,\,  \sum_{j=0}^{\infty}\,b_{2n+1,j}\cdot \,\tau^{j}.
\end{eqnarray}
The functions $g_t,~g_h$ may depend on the interaction energies
$E(j,k)$ in (\ref{geni}), but the constant ${\tilde Y}$, and the
functions $Y_{\pm}(z)$, do not depend on the interaction energies.
The functions $Y_{\pm}(z)$ are referred to as  ``scaling functions''.
 The independence on the constants $E(j,k)$ is 
referred to as ``universality''.
The leading behavior as $T\,\rightarrow\, T_c$,
 and $H\,\rightarrow \,0$, of 
$\, g_h/|g_t|^{15/8}$ is:
\begin{equation}
g_h/|g_t|^{15/8}\,\,\,\, \sim \,\,\,\,
\frac{b_{1.0}H}{|a_{0,1}\,\tau|^{15/8}}\,
\,\, =\,\,  h, 
\label{tscaling}
\end{equation}
(where the last line is a definition).
When $H\neq 0$ the free energy is analytic for all real values of $T$,
and, therefore, the functions $Y_{+}(z)$ and $Y_{-}(z)$ must analytically
connect with each other as $z\rightarrow  \infty$. These are, in fact,
parts of the same function. This is conveniently expressed by defining
a new variable 
\begin{equation}
\eta\,\, =\,\,\,\,\,  \lim_{H\rightarrow 0,\tau\rightarrow 0}\,\frac{\tau}{|H|^{8/15}}. 
\label{etadef}
\end{equation}
Instead (\ref{af1}), we follow \cite{fz} in 
defining the ``singular'' part of scaling free
energy, for $\tau$ and $H$ real, as
\begin{equation}
F\,\, =\,\,\,  \frac{\tau^2}{8\pi}\cdot \, \ln \tau^2\,\, \, 
+|H|^{16/15}\cdot \, \Phi(\eta),
\end{equation}
(where we have used the normalization conventions of~\cite{fz} with
the exception that we use $-\tau$ instead of $m$). 
For $H$ and $\tau$ both real and positive, we have
\begin{equation}
\eta \,\, =\,\,\, \, 1/h^{8/15}.
\end{equation}
Thus, for real $\eta>0$ ($T>T_c$), we have
\begin{equation}
\Phi(\eta)\,\, =\,\,\,  \, \eta^2 \cdot \, Y_{+}(1/\eta^{15/8}). 
\label{phip}
\end{equation}
For $H>0$ and $\tau<0$ ($T<T_c$), we have 
\begin{equation}
\eta\,\,  =\,\,\, \,   -1/h^{15/8}.
\end{equation}
Thus, for real $\eta<0$ we have:
\begin{equation}
\Phi(\eta)\, =\, \,\,  \, \eta^2 \cdot \,Y_{-}(1/(-\eta)^{15/8}). 
\label{phim}
\end{equation}

The function $\Phi(\eta)$
has, for large values of $\eta$ on the
real line of~\cite{fz}, the behavior 
\begin{eqnarray}
\hspace{-0.3in}&&\Phi(\eta)\,\,=\,\,\,\, 
\eta^2 \cdot \,\sum_{k=1}^{\infty}\,{\tilde G}_k \cdot \, (-\eta)^{-15k/8},
\quad \quad \quad \quad \eta\, \rightarrow \,-\infty,  \\
\hspace{-0.3in}&&\Phi(\eta)\,\,=\,\,\,\, \eta^2 \cdot \, \sum_{k=1}^{\infty}
G_k \cdot \, \eta^{-30k/8},\quad \quad \quad \quad 
\eta \, \rightarrow \, \infty,
\end{eqnarray}
and, for small values of $\eta$:
\begin{equation}
\Phi(\eta)\,\,=\, \,\,\, -\frac{\eta^2}{8\pi}\,\ln \eta^2\,\,\,
+\sum_{k=0}^{\infty}\,\Phi_k\cdot \, \eta^k.
\end{equation}

For real values of $\eta$ the function
$\Phi(\eta)$ has been numerically determined~\cite{man}
 by Baxter's variational
approach, based on the corner transfer matrix~\cite{bax2},~\cite{bax3},
as enhanced by an improved iteration scheme~\cite{nishino}. 

From the form (\ref{af1})  Aharony and Fisher~\cite{af} derive
an expression for the functions $F_{\pm}(\tau)$ of (\ref{series}),
which is supposed to have validity, not only in the limit
$H\, \rightarrow \, 0,~\tau \, \rightarrow \,0$,
 but also for $H$ and $\tau$ finite.
Unfortunately, the conjectured form (\ref{af1}) fails for several reasons.
First of all, the result of~\cite{af} has $F_{+}(\tau)=\, F_{-}(\tau)$,
whereas the analysis of~\cite{ongp}, and~\cite{cgnp}, show 
that $F_{\pm}(\tau)$ differ, at order $\tau^6$, on the square, triangular
and hexagonal lattice. In addition, for the square lattice, the term
$\tau^4$ of~\cite{af} disagrees with~\cite{ongp},~\cite{cgnp}.
Furthermore the conjecture (\ref{af1}) does not account for the term
$B(\tau)$ of (\ref{series}). Clearly a more general conjecture is required.

A more general conjecture, than that of Aharony and Fisher, utilizes the
``irrelevant variable'' of scaling theory. We  follow
the notation of~\cite{ongp} and~\cite{cgnp}. We write this more
general conjecture, for the singular part of the free energy, as
\begin{eqnarray}
\hspace{-0.5in}&&f_{\rm sing}(g_t,g_h,\{g_{u_j}\})
\,\,=\,\,\, \, g^2_t \,\ln |g_t|\cdot 
{\tilde Y}_{\pm}(g_h/|g_t|^{15/8},\{g_{u_j}/|g_t|^{y_j/y_t}\})
\nonumber\\
\hspace{-0.5in}&&\qquad \qquad \qquad \qquad 
+g^2_t\cdot Y_{\pm}(g_h/|g_t|^{15/8},\{g_{u_j}/|g_t|^{y_j/y_t}\}),
\label{af2}
\end{eqnarray}
where, now, the coefficients $a_{2n}(t)$ and $b_{2n+1}$, in the nonlinear
scaling fields in (\ref{nonlinear}), are allowed to depend on a set of
variables $u_j$,
and where
\begin{equation}
g_{u_j}\,\,\, =\, \,\,\,\, \sum_{n=0}\,c_{2n}^j(t,u) \cdot \,H^{2n},
\end{equation} 
are additional ``nonlinear scaling fields'' associated with 
the irrelevant operators of scaling theory that have dimensions
$y_j$. Further definitions of these concepts may be found for example
in \cite{caselle}. However, no explicit forms, or conjectures,
for these multivariable formulas have ever been given. Moreover 
there is no prescription given to separate the effects of the 
``non linear scaling fields'' from the irrelevant operators. 
In addition the higher powers of $\ln \tau$, seen in the 
susceptibility, are not present in the form (\ref{af2}). Consequently,
while the form (\ref{af2}) may be regarded as ``conventional'', it 
is descriptive rather than computational. Indeed, in~\cite{ongp}, 
and~\cite{cgnp}, even though it is introduced, it is  never used.

A further difficulty with the ``scaling form'' (\ref{af2}) is that it
does not give an explanation for the higher powers of $\ln |\tau|$
which occur in $B(\tau)$ of (\ref{short}). At this point the literature
is slightly ambiguous. In~\cite{ongp}, in footnote 12, it is stated that
the fact that there is a ``resonance between the identity, and the
energy'', will result in higher powers of $\ln \tau$ (much as integer
differences of exponents lead to powers of logarithms in Fuchsian
differential equations). However, in footnote 5 of~\cite{cgnp}, 
it is said that these
powers of $\ln |\tau|$ have ``not yet been interpreted within the
context of scaling theory''. There are clearly things left to be
understood.

To see what is needed, we generalize our point of view, from the nearest
neighbor Ising model (\ref{nni}), to the much more general case
(\ref{geni}), which allows for many further neighbor interactions. 
This more general model is not integrable. Now the internal energy  
(\ref{genu}) will include the spin correlations of all spins which are
connected by non zero interaction energies. Thus, at least for small
values of the non planar bonds, it is entirely reasonable to expect that,
at $T\rightarrow T_c$, the singularity will not be a pure logarithm, as
is the case for the nearest neighbor case, but will involve many higher
powers of $\ln^p |\tau|$, each of which are multiplied by some suitably
high power of $\tau$. We might speculate, from (\ref{short}), that this power 
will be $\tau^{p^2}$, but  there is no further argument for this,
except that it is the result found in (\ref{short}). Moreover, even
in the case of the nearest neighbor interactions (\ref{nni}), these
terms, with higher powers of $\ln |\tau|$, should be present.  
By including such terms we will be able to reproduce the ``short range''
term $B(\tau)$ in the susceptibility expansion (\ref{chishort}).
 
But the truly serious problem in obtaining a general form of the free energy
of the Ising model in a magnetic field, as an expansion for ``small $H$
and small $\tau$'', is that the natural boundary, discussed in section 3 for
the susceptibility, forces the expansion (\ref{series}) to be
asymptotic. Therefore, even though the susceptibility is well defined
for all values of $\tau$, the expansion (\ref{series}) cannot be taken
as the definition of the susceptibility.  This is exactly the same
problem which afflicts perturbation expansions in the Feynman diagram
expansion of quantum field theories, such as Quantum 
Electrodynamics~\cite{dyson},~\cite{thooft}.

\subsection{Non-integrable perturbations for $H\neq 0,~~ T\neq 0$}

The most ambitious program of studying the Ising model, for $H\neq \,0$,
is to allow both $T\neq \,T_c$, and $H\neq 0$, at the same time. This will
allow the passage, from $T<\,T_c,~H=\, 0$ to $T>T_c$, $ \,H=\,0$, on a path in the
real $(H,T)$ plane, as is contemplated in the scenario of~\cite{mccoy52}.  
However this more general two variables perturbation does not satisfy
the criteria of~\cite{zam},~\cite{zam2}
 needed for an integrable perturbation.
Furthermore, in this more general case, we
cannot treat the region of $H\sim\, 0$, and $T\sim\, T_c$, in isolation
from the Lee-Yang edge for complex $H$. Thus, we need to
understand how the $M(3,4)$ conformal field theory flows to the
$M(2,5)$ theory under the influence of the nonintegrable perturbation.

To make this precise it is necessary to define a scaling limit in 
the complex plane. This is very delicate 
because it says that we are describing the singular behaviour of a function of
two complex variables in terms of one complex variable. Such a
reduction will require very special circumstances to be valid. 

Some of the consequences  of the existence of this limit have for the
free energy are extensively discussed in the 2003 
paper of Fonseca and Zamolodchikov~\cite{fz}, where  
a scenario is presented which incorporates several assumptions about
analytical continuation in $T$ and $H$,
 as two complex variables. We note, however, that the question of the
relation of the natural boundary in the complex $T$ plane to the
analyticity properties in the complex H plane near the Lee-Yang edge
remains to be investigated.

\section{Why the Ising model is important}

It is very natural to extend perturbed conformal field theory 
of the Ising model from the integrable cases 
of~\cite{zam},~\cite{zam2}, where 
either $H=\,0,~T\,\neq\, T_c$, and $T=\, T_c, H\neq 0$, to the general case
$T \,\neq \,T_c, \,H\,\neq \,0$. This has been done
 in~\cite{fz}. However, in
spite of the impressive results of~\cite{man} for the free energy,
which are in correspondence with the picture of~\cite{fz}, 
there are several very interesting questions which remain to be
addressed.

Perhaps the foremost of these questions is what may be a significant
difference between integrable and non-integrable systems. We saw, in
section 4.4, that the characters of the conformal field theory of the
$M(3,4)$ minimal model have three different representations which
are in one to one agreement with the spectrum of excitations, in the
massive case, away from the critical point. We can thus think of 
 the integrable perturbations as being precisely tuned to these
three different bases, in the same way as degenerate perturbation theory
picks out a distinguished basis. 

However, in the non-integrable cases, the spectrum of excitations
contains a number of particles which depends on both $T$ and $H$, and
is not just either one or eight. 
The implication of this mismatch between the variable number of excitations,
seen in the massive model (in what can be considered the infrared part
of the spectrum), and the  existence of only three particle-like
representations of the character in the conformal field theory (which
can be considered as the ultraviolet part of the spectrum), is that
the smooth match, found in the Ising model at $H=0$ (which was 
shown in section 3.8 with the precise demonstration by
 Tracy~\cite{tracy2} of the equality (\ref{equality})),
 may not hold in the general non-integrable case.

The discussion of the potential disagreement of the short distance
behavior of the scaled two-point correlation function for general
values of $T$ and $H$ of the Ising model, with the result expected from
conformal field theory, raises the suggestion that our understanding of
scaling theory may not be complete. 

Scaling theory is concerned with the relation of two length
scales: the scale of the lattice length on which the theory is defined,
and the correlation length which is observed in the correlation
functions. In scaling theory, as used in statistical mechanics, 
and in renormalization of quantum field theory, there is a smooth match 
between these two scales. But this is drastically different from what
occurs in systems, such as fluid mechanics, where
 a common piece of folk wisdom of perturbation theory~\cite{boyd} 
is  that ``divergence should
be expected when the solution depends on two independent length
scales''. This  would
also seem to be related to the fact, found in~\cite{ongp} and~\cite{cgnp},
 discussed in section 3.8, that the susceptibility does not
have a convergent expansion about $T=\,T_c$. It would also seem to be
 in agreement with the fact
that the natural boundary of the susceptibility, presented in section
3.7, does not naturally fit into conventional scaling theory. 
Thus it may be the case that the smooth matching of long and short
distance expansions of the Ising model, at $H=\, 0$, is caused by the
integrability of the model, and may not hold in the general 
non-integrable case.

 For these reasons it can
certainly be said that, even though the Ising model is the best
understood system in statistical mechanics, there are still many
puzzling questions to be investigated, questions which have an
importance well beyond the narrow range of just this one model. In
fact, it can be argued that many of the nagging questions concerning
our understanding of quantum field theory~\cite{dyson},~\cite{thooft}
are related to the puzzles of the Ising model. Hopefully these
questions will inspire future research.

\vspace{.2in}

\centerline{\bf \Large Acknowledgments}

\vspace{.1in}

It is a great pleasure to thank  J.H.H. Perk for many insightful
discussions and S. Hassani for unpublished results. 
We also thank C. M. Bender, A. J. Guttmann, B. Nickel, 
C. A. Tracy, F. Y. Wu for useful comments. One of us (BMM) is grateful
to T. Deguchi for the opportunity of presenting a preliminary version
this review at the ``Tokyo Mathematical Physics Seminar'' in July 2011.

\end{document}